\documentclass[letter]{aa}
\usepackage{graphicx}
\usepackage{txfonts}
\usepackage{xcolor}
\begin{document} 

   \title{A cautionary tale in fitting galaxy rotation curves with Bayesian techniques: does Newton's constant vary from galaxy to galaxy?}
   \titlerunning{Does Newton's constant vary from galaxy to galaxy?}

   \author{Pengfei Li
          \inst{1}\fnmsep\thanks{Email: PengfeiLi0606@gmail.com, pxl283@case.edu}
          \and
          Federico Lelli\inst{2}
          \and
          Stacy McGaugh\inst{1}
          \and
          James Schombert\inst{3}
          \and
          Kyu-Hyun Chae\inst{4}
          }

   \institute{Department of Astronomy, Case Western Reserve University, Cleveland, OH 44106, USA
         \and
             School of Physics and Astronomy, Cardiff University, Queens Buildings, The Parade, Cardiff, CF24 3AA, UK
             \and Department of Physics, University of Oregon, Eugene, OR 97403, USA
             \and Department of Physics and Astronomy, Sejong University, 209 Neungdong-ro Gwangjin-gu, Seoul 05006, Republic of Korea
             }

   \date{Received xxx; accepted xxx}
 
  \abstract
{The application of Bayesian techniques to astronomical data is generally non-trivial because the fitting parameters can be strongly degenerated and the formal uncertainties are themselves uncertain. An example is provided by the contradictory claims over the presence or absence of a universal acceleration scale (g$_\dagger$) in galaxies based on Bayesian fits to rotation curves. To illustrate the situation, we present an analysis in which the Newtonian gravitational constant $G_N$ is allowed to vary from galaxy to galaxy when fitting rotation curves from the SPARC database, in analogy to $g_{\dagger}$ in the recently debated Bayesian analyses. When imposing flat priors on $G_N$, we obtain a wide distribution of $G_N$ which, taken at face value, would rule out $G_N$ as a universal constant with high statistical confidence. However, imposing an empirically motivated log-normal prior returns a virtually constant $G_N$ with no sacrifice in fit quality. This implies that the inference of a variable $G_N$ (or g$_{\dagger}$) is the result of the combined effect of parameter degeneracies and unavoidable uncertainties in the error model. When these effects are taken into account, the SPARC data are consistent with a constant $G_{\rm N}$ (and constant $g_\dagger$).
}


   \maketitle

\section{Introduction}

Bayesian analysis has been extensively implemented in astronomical studies due to the frequent occurrence of multi-parameter problems. A major philosophy of Bayesian inference is to incorporate reliable priors to break the parameter degeneracies. This is best achieved by imposing physically motivated priors which, however, are not always readily available. The situation becomes even more difficult when the input formal uncertainties are themselves uncertain, as is often the case for heterogeneous collections of data. Thus, any formal outcome of Bayesian analysis should be interpreted with a proper appreciation for both the strengths and limitations of the source data.

One example is the claim of ``absence of a fundamental acceleration scale in galaxies" by \citet{Rodrigues2018} and the follow-up work by \citet{Marra2020}. They fitted rotation curves of disk galaxies from the SPARC\footnote{Spitzer Photometry and Accurate Rotation Curves. All data are available at \url{astroweb.case.edu/SPARC}.} database \citep{SPARC} using the radial acceleration relation \citep[RAR,][]{McGaugh2016PRL, OneLaw}. The RAR is a tight empirical relation between the observed kinematic acceleration g$_{\rm obs}$ and the baryonic gravitational field g$_{\rm bar}$ with a characteristic scale g$_\dagger \simeq 10^{-10}$ m s$^{-2}$. Its asymptotic behaviors agree with the predictions of modified Newtonian dynamics \citep[MOND,][]{Milgrom1983}, hence the RAR can be interpreted as a viable realization of MOND only if the empirical scale g$_\dagger$ is a universal constant.

\citet{Rodrigues2018} tested the universality of g$_\dagger$ by fitting the RAR to individual galaxies imposing flat priors on three fitting parameters (acceleration scale g$_\dagger$, galaxy distance, and stellar mass-to-light ratio) and taking formal uncertainties literally for each and every rotation curve. They obtained a wide distribution of g$_\dagger$ and concluded that a universal acceleration scale is ruled out. After substantial criticism about their methodology and interpretation of data \citep{McGaugh2018, Kroupa2018, Cameron2020}, they presented a follow-up analysis in \citet{Marra2020}, in which they follow \citet{Li2018} in adopting physically motivated Gaussian priors on distance, inclination, and stellar mass-to-light ratio. They persist in using a flat prior on g$_{\dagger}$, resulting in severe parameter degeneracies. Similar work was presented by \citet{Chang2019} using a very broad, uninformative prior on g$_\dagger$.
 
The mere existence and tightness of the RAR and of the baryonic Tully-Fisher relation \citep[BTFR, e.g.,][]{Lelli2019} suggest an empirically motivated Gaussian prior centered around 10$^{-10}$ m s$^{-2}$. There cannot be more scatter in $g_\dagger$ than there is in the raw forms of these relations prior to any rotation-curve fitting, yet this is exactly what is inferred in analyses where a broad prior on $g_\dagger$ is assumed. Degeneracy can cause this parameter to vary widely with other parameters without providing a meaningful improvement in fit quality. This was already pointed out in \citet{Li2018}, where we concluded that rotationally supported galaxies are consistent with a single value of g$_\dagger$: there is no value added in allowing it to vary. The same argument applies to pressure supported galaxies \citep{Chae2020ApJL} and when the MOND external-field effect is taken into account \citep{Chae2020}.

Comparing the posterior distribution functions of $g_\dagger$ from individual galaxy fits implies that one can fully trust the formal uncertainties in each and every case. The observational uncertainties of SPARC galaxies are surely sensible on average; for example, the mean expected scatter in the RAR from error propagation is comparable with the observed scatter \cite[e.g.,][]{OneLaw}. However, taking formal uncertainties of each and every rotation curve literally is not recommended because they were culled from multiple sources \citep{SPARC}. More generally, we can measure the Doppler shift in different parts of galaxies with high accuracy, but dynamical analysis demands knowledge of the circular velocity of a test particle in the equilibrium gravitational potential. The gas rotational velocity is arguably the best possible proxy to the circular velocity, but uncertainties in some individual galaxies (e.g., due to non-circular motions or out-of-equilibrium dynamics) can never be fully under control. 

To further illustrate how a blind application of Bayesian statistics with uninformative priors could mislead us, we present a ``reductio ad absurdum'' using Newton's gravitational constant $G_N$ as a free parameter in a similar fashion as g$_\dagger$. Choosing $G_N$ has a key advantage: since it is widely accepted as a constant, comparing the Bayesian inference with the expected value provides a clear evaluation on the approach of \citet{Rodrigues2018} with respect to that of \citet{Li2018}. In Section 2, we reproduce the results by \citet{Rodrigues2018} using similar data and priors but for $G_N$ instead of g$_\dagger$. In Section 3, we repeat the analysis of \citet{Li2018} exploring different priors on $G_{N}$. We discuss our results in Section 4.

\section{Challenging the constancy of Newtonian gravitational constant}

In analogy to previous works \citep{Li2018, Rodrigues2018, Chang2019, Marra2020}, we fit the rotation curves from the SPARC database \citep{SPARC} using the RAR. The SPARC database consists of 175 late-type galaxies with accurate H\,{\footnotesize I}/H$\alpha$ rotation curves and mass modeling from Spitzer photometry at 3.6 $\mu$m. The gravitational contributions of different baryonic components (gas disk, stellar disk, and bulge if applicable) are represented as the circular velocity of a test particle ($V_{\rm gas}$, $V_{\rm disk}$, and $V_{\rm bulge}$, respectively). The total rotation velocity due to baryonic contributions $V_{\rm bar}$ is then the quadratic sum of these components:
\begin{equation}
    V_{\rm bar}^2 = V_{\rm gas}|V_{\rm gas}| + \Upsilon_{\rm disk}V_{\rm disk}^2 + \Upsilon_{\rm bulge}V_{\rm bulge}^2,
\end{equation}
where $\Upsilon_{\rm disk}$ and $\Upsilon_{\rm bulge}$ are the stellar mass-to-light ratios for disks and bulges with the fiducial values $\Upsilon_{\rm disk}=0.5$ and $\Upsilon_{\rm bulge}=0.7$ \citep{McGaugh2016PRL}. The RAR relates the observed acceleration g$_{\rm obs}$ to that due to baryons g$_{\rm bar}$:
\begin{equation}\label{eq:RAR}
    {\rm g_{obs}} = \frac{\rm g_{bar}}{1-e^{-\sqrt{\rm g_{bar}/g_\dagger}}},
\end{equation}
where ${\rm g_{obs}} = {\rm V_{obs}^2}/R$, ${\rm g_{bar}} = {\rm V_{bar}^2}/R$, and g$_\dagger$ is the only free parameter. This relation was established by analyzing the data points from 153 galaxies in a statistical sense. We aim to test the constancy of $G_N$, so we fit Eq.\,\ref{eq:RAR} to rotation curves fixing $\rm{g_\dagger= 1.2\times 10^{-10}}$ m\,s$^{-2}$ and varying $G_N$ from galaxy to galaxy. We also tried to vary g$_\dagger$ and $G_N$ simultaneously,  obtaining similar results. We show as an example the case that has the same number of fitting parameters as in \citet{Rodrigues2018}.

\begin{figure}
    \centering
    \includegraphics[scale=0.4]{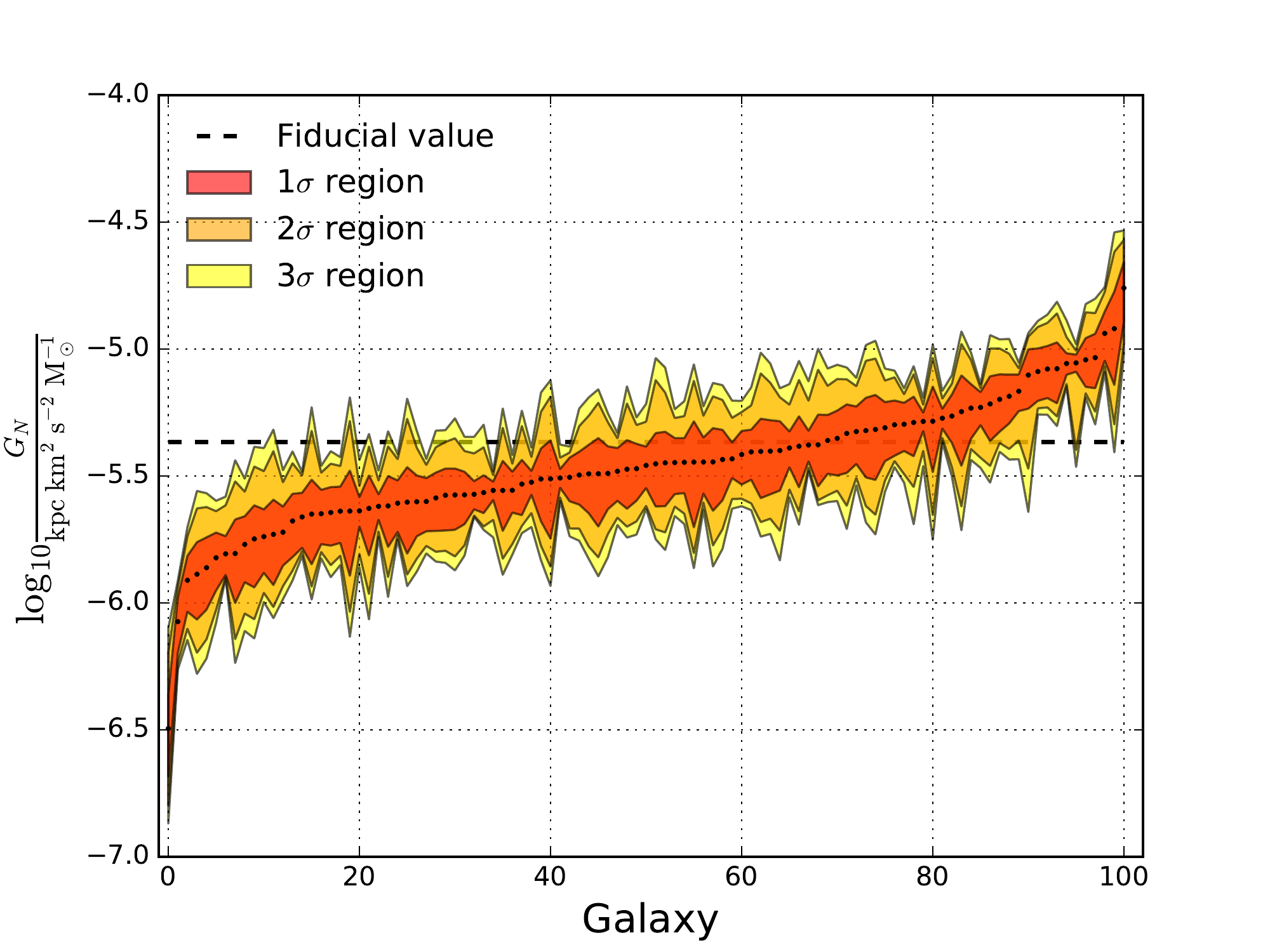}
    \caption{Posterior probability distribution of Newton's gravitational constant $G_N$ for the galaxies in the SPARC database \citep{SPARC}. Following a similar analysis by \citet{Rodrigues2018} about the acceleration scale $g_\dagger$, galaxies are ordered for increasing value of $G_N$ and flat priors are imposed on galaxy distances, stellar mass-to-light ratios, and $G_N$ (see text for details). Black dots show the maximum of the posterior probability, together with 1$\sigma$, 2$\sigma$ and 3$\sigma$ credible regions from the Python package GetDist. The expected value of $G_N$ is represented by the dashed line. Among the selected 101 galaxies, 34 are incompatible with the expected $G_N$ at the 3$\sigma$ level. Taken at the face value, this would imply that the gravitational constant is not constant.}
    \label{fig:GN}
\end{figure}

We set the likelihood function as $L=e^{-\frac{1}{2}\chi^2}$ with $\chi^2$ given by
\begin{equation}
    \chi^2=\sum_R\frac{{\rm [g_{obs}(R)-g_{RAR}(R)]^2}}{\rm\delta g_{obs}^2(R)},
\end{equation}
where ${\rm g_{RAR}}$ is the expected centripetal acceleration from Eq.\,\ref{eq:RAR} and ${\rm\delta g_{obs} = 2V_{obs} \frac{\delta V_{obs}}{R}}$ is the uncertainty on the observed acceleration. We impose a flat prior on $G_N$ with $G_N>0$ as \citet{Rodrigues2018} did for g$_\dagger$. Following the procedure of \citet{Rodrigues2018}, we also impose flat priors on $\Upsilon_\star$ with a tolerance of a factor of two, though this is not consistent with stellar population synthesis models \citep{Schombert2019}.

Galaxy distance affects the fit quality as it is directly related to the contributions of the baryonic components. When the distance $D$ is changed to $D'$, the galaxy radius $R$ and the contribution of each component $V'_k$ will transform according to
\begin{equation}
    R'=R\frac{D'}{D}, \quad V'_k = V_k\sqrt{\frac{D'}{D}},
\end{equation}
where $k$ denotes gas, disk, or bulge. \citet{Rodrigues2018} allow galaxy distance to vary freely within 20\% of their observational values. This prior does not reflect actual observational errors. Distances measured through Cepheids or the tip magnitude of the red giant branch are highly accurate, so a 20\% variation could be larger than 4$\sigma$. The Hubble flow method is much less accurate; distances measured using this method could have errors up to 30\%, so the imposed 20\% free range is even within the 1$\sigma$ region. \citet{Rodrigues2018} also ignored the uncertainty on disk inclination. Possible outer asymmetries and warps in the gas disk could mislead the determination of inclinations, as reflected in their uncertainties, which would in turn affect the observed rotation velocities. Despite the shortcomings of the methodology of \citet{Rodrigues2018}, we choose to fit the same parameters (except $G_N$ instead of $g_\dagger$) and impose the same priors to achieve the most direct comparison. 

\begin{figure*}
    \centering
    \includegraphics[scale=0.43]{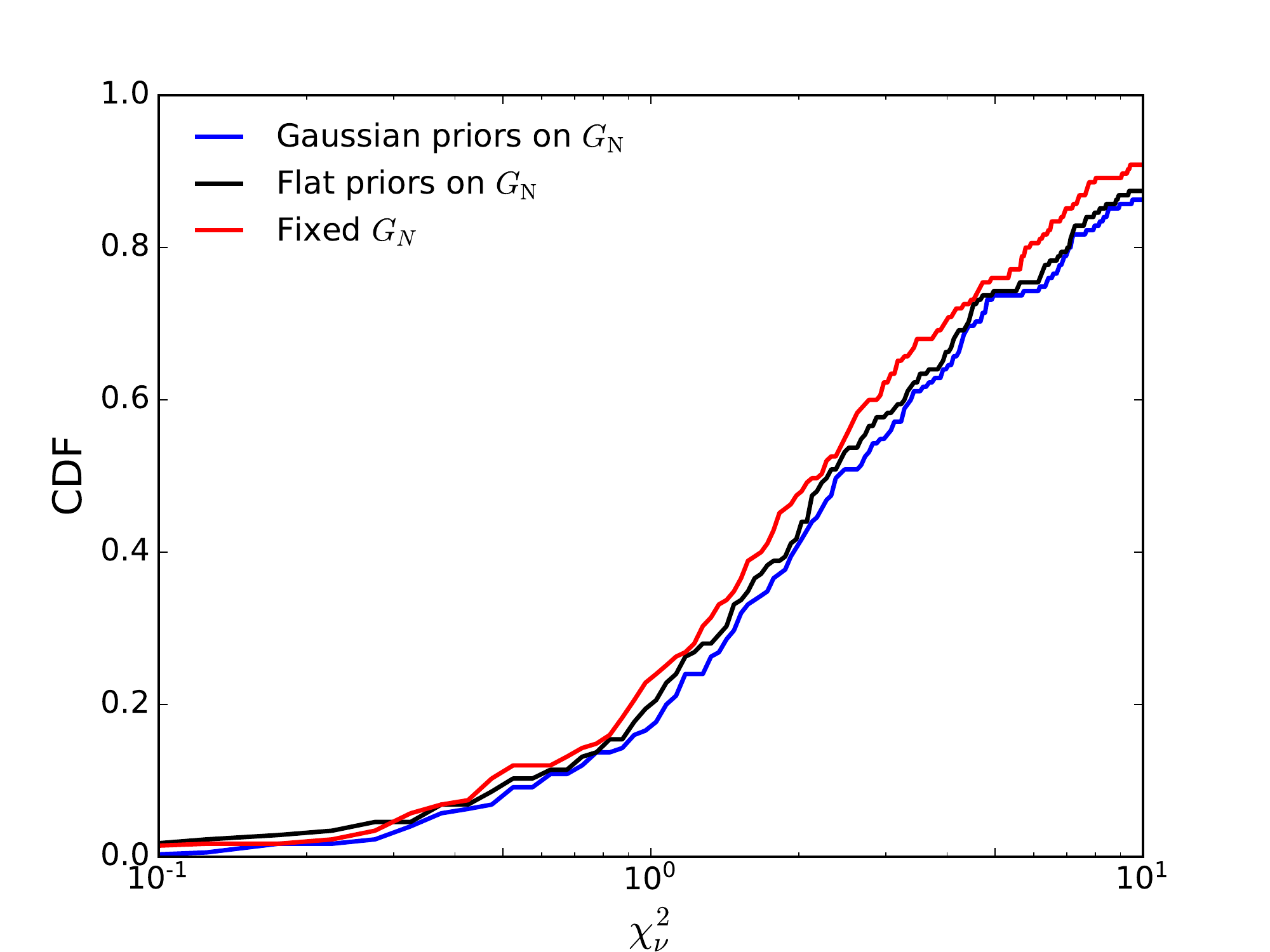}\includegraphics[scale=0.43]{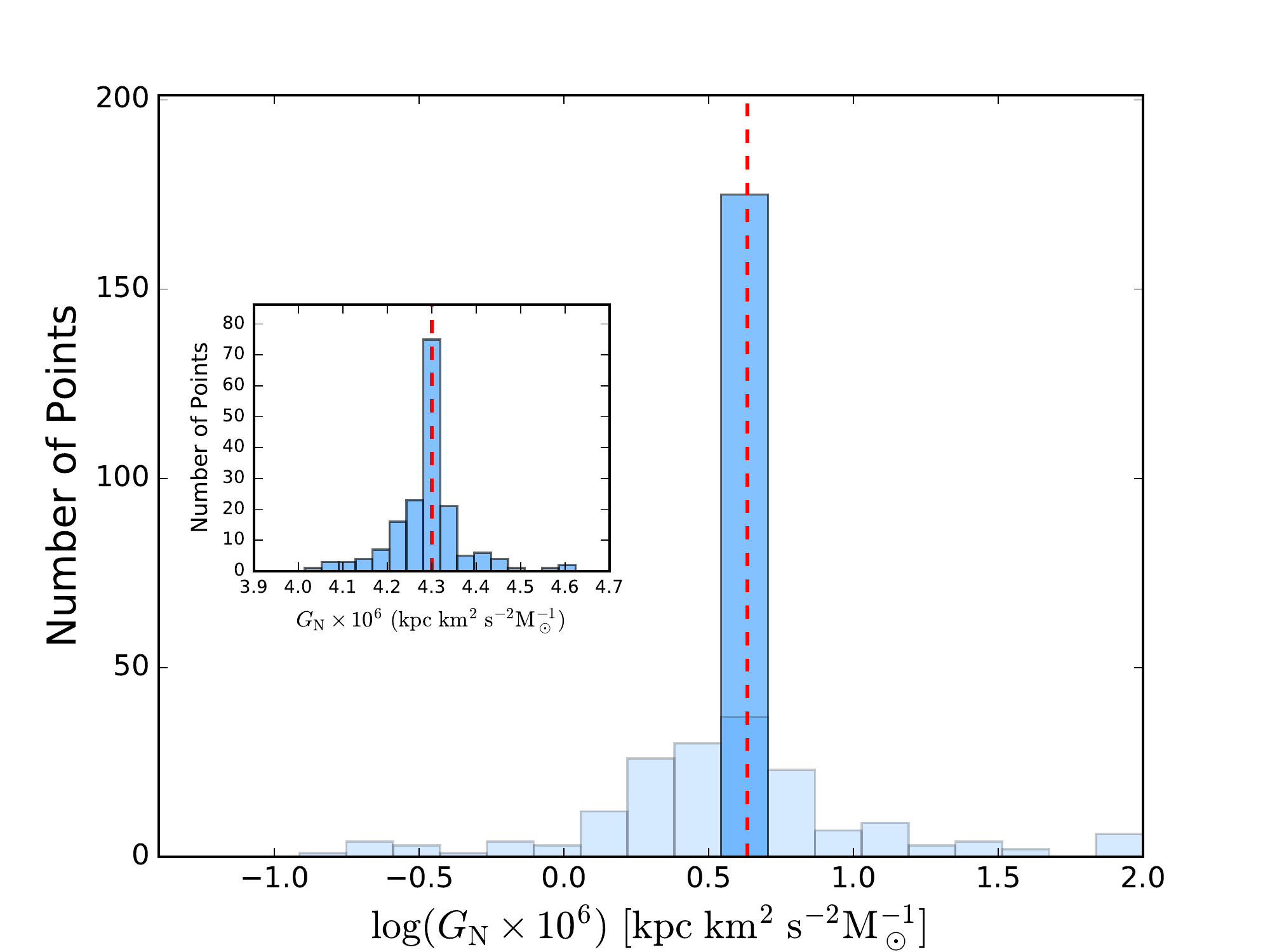}
    \caption{Left: cumulative distribution functions of $\chi^2_\nu$ fixing $G_N$ (red line) or imposing empirical lognormal priors (blue line) and flat priors (black line) on $G_N$. The three cases show indistinguishable fit qualities. Right: histograms of the maximum-likelihood $G_N$. The dark and light blue histograms correspond to lognormal and flat priors on $G_N$, respectively. The inset panel shows a zoom-in distribution for the Gaussian prior, switched to linear scale for a better view. The actual value of $G_N$ is indicated by red, dashed lines. The Gaussian prior returns a tight distribution of $G_N$ without decreasing the fit quality: this confirms the constancy of $G_N$ as expected. The same argument has been applied for the constancy of g$_\dagger$ in \citet{Li2018}. }
    \label{fig:CDFGN}
\end{figure*}

We map the posterior distributions of the fitting parameters using the standard affine-invariant ensemble sampler in the Python package $emcee$ \citep{emcee2013}. The Markov Chains are initialized with 200 random walkers and iterated for 2000 steps after 1000 burnt-in iterations. We record the best-fit parameters that maximize the probability function for every galaxy.

Figure \ref{fig:GN} shows the posterior probability distribution of $G_N$. Though we analyze all the SPARC galaxies, we only include 101 of them in Figure \ref{fig:GN} based on the following quality cuts: we removed galaxies with quality flag Q=3 \citep[see][]{SPARC}, inclination smaller than 30$^{\circ}$, or reduced $\chi^2_\nu>4.0$. This is a similar quality cut as adopted in \citet{Rodrigues2018}, and it retains a similar number of galaxies. Figure \ref{fig:GN} shows that the posterior probability distribution of $G_N$ presents a wide distribution spanning $\sim$3.5 dex. We also calculate the 1$\sigma$, 2$\sigma$, and 3$\sigma$ credible regions (including 68.3\%, 95.4\% and 99.7\% of the posterior probability, respectively) from the output of ``getMargeStats" in the Python package GetDist\footnote{\url{https://getdist.readthedocs.io}}. It turns out that 34 galaxies out of 101 are incompatible with the expected $G_N$. This is similar to Figure 1 of \citet{Rodrigues2018}, showing a wide distribution of $a_0$ (which we call g$_\dagger$ to distinguish empirical free parameters from physical constants). \citet{Rodrigues2018} found that 31\% of the galaxies are rejected from the global best-fit g$_\dagger$ at 3$\sigma$ level. Hence, they infer the ``absence of a fundamental acceleration scale in galaxies''. We find 34\% of the galaxies are excluded from the expected value of $G_N$ at 3$\sigma$ level. Following their argument, we reach the conclusion of the absence of a fundamental Newtonian gravitational constant, ruling out Newtonian gravity and General Relativity.

There are two interconnected problems here: (i) flat priors can give $G_N$ too much freedom to go astray, and (ii) formal uncertainties on circular velocities (as well as distances and inclinations) are themselves uncertain, so the confidence regions from the posterior distribution functions of individual galaxies have to be taken with a grain of salt. With highly accurate observational data (e.g.\ the orbital data of the planets in the solar system), one would expect that flat priors would return consistent values of $G_N$. For observational data like galaxy rotation curves, there can occur cases where parameters co-vary wildly given the imperfect nature of the input data and their uncertainties (e.g. warped disks, complex non-circular motions, etc.). Since flat prior does not provide any constraints against these effects, the fitting parameters can go astray out of reasonable regions. 

\section{Confirming the constancy of Newtonian gravitational constant}

To further demonstrate the degeneracy problem, we compare fit qualities imposing different priors on $G_N$. Similar work has been done in \citet{Li2018} for g$_\dagger$. In that paper, we found that imposing flat and Gaussian priors on g$_\dagger$ result in significantly different distributions of its maximum likelihood value, but essentially return the same fit quality. We concluded that adding g$_\dagger$ as a fitting parameter does not improve fit quality, thus the data are consistent with a single value of g$_\dagger$. The validity of this argument should be straightforward, but given the recent claims of \citet{Chang2019} and \citet{Marra2020}, we repeat the analysis of \citet{Li2018} varying $G_N$.

In \citet{Li2018}, we include disk inclination as a fitting parameter as it affects the observational rotation velocities and their uncertainties. When the inclination is adjusted from $i$ to $i'$, the observational rotation velocities and the uncertainties will transform as
\begin{equation}
    V'_{\rm obs} = V_{\rm obs}\frac{\sin{i}}{\sin{i'}}; \quad \delta V'_{\rm obs} = \delta V_{\rm obs}\frac{\sin{i}}{\sin{i'}},
\end{equation}
where $V_{\rm obs}$ and $\delta V_{\rm obs}$ are observational velocities and uncertainties, respectively. We also impose more realistic priors than \citet{Rodrigues2018} on the fitting parameters: a lognormal prior on the stellar mass-to-light ratio with a standard deviation of 0.1 dex inspired by stellar population synthesis model \citep[e.g.,][]{Schombert2019}, and Gaussian priors on galaxy distance and disk inclination with the standard deviations given by their observational uncertainties. The same priors on these nuisance, galactic parameters are imposed here.

For $G_N$ we follow two different approaches: (1) we impose an uninformative flat prior within [10$^{-8}$, 10$^{-4}$] kpc km$^2$ s$^{-2}$ M$_\odot^{-1}$ which is known to give unrealiable results (as shown in the previous section), and (2) we use a lognormal prior following the so-called ``empirical Bayes approach''. The lognormal prior is centered around the expected value $G_N = 4.3\times 10^{-6}$ kpc km$^2$ s$^{-2}$ M$_\odot^{-1}$, while its standard deviation is empirically motivated using the BTFR \citep{Lelli2019}. The BTFR is mathematically equivalent to the low-acceleration portion of the RAR \citep[see Sect. 7.1 in][]{OneLaw} and can be expressed as:
\begin{equation}\label{eq:BTFR}
\log(V_{\rm f}) = 0.25\log(M_{\rm b}) + 0.25\log(g_\dagger G_N /X),
\end{equation}
where $V_{\rm f}$ is the flat rotation velocity, $M_{\rm b}$ is the total baryonic mass, and $X$ is a factor of order unity that accounts for the cylindrical geometry of disks \citep{McGaugh2018}. Fitting the data from \citet{Lelli2019} in log-log space with LTS\_LINEFIT \citep{Cappellari2013}, we find that the best-fit slope is indistinguishable from 0.25 and the vertical intrinsic scatter is 0.02 dex along $V_{\rm f}$. This sets a very hard upper limit for the scatter on $G_N$ or $g_\dagger$ given that galaxy-to-galaxy variations in $X$ may also contribute. Any intrinsic variation in $\log(G_{\rm N})$ or $\log(g_\dagger)$ just cannot be larger than 0.02 dex. Thus, we consider a standard deviation of 0.02 dex for the lognormal prior on $G_{\rm N}$. We stress that $G_{\rm N}$ and $g_\dagger$ enter in Eq.\,\ref{eq:BTFR} in the same fashion, so the same argument can be applied to $g_\dagger$ analogously to \citet{Li2018}.

The left panel of Figure \ref{fig:CDFGN} shows the cumulative distribution function (CDF) of the reduced $\chi^2$ for both flat and lognormal priors on $G_N$. We stress that $\chi^2_{\nu}$ is merely used as a first-order statistics to assess the overall quality of different Bayesian fits, so the arguments against its use outlined in \citet{Rodrigues2018b} do not apply. Moreover, we are not comparing individual galaxy fits, but their average, cumulative behavior that is more robust against the occasional overestimate or underestimate of formal uncertainties. Unsurprisingly, the results shown here are similar to Figures 6 and 7 in \citet{Li2018}, though we are varying $G_N$ rather than g$_\dagger$. The CDF shows that imposing flat and lognormal priors essentially give the same fit quality. For reference, we also include the results with fixed $G_N$ (red solid line), which shows a slightly better CDF of $\chi^2_\nu$ despite it has one less fitting parameter. This occurs because the rotation-velocity residuals are comparable, so the smaller number of fitting parameters $f$ for the same number $N$ of data points improves the value of $\chi^2_{\nu} = \chi^2/(N-f)$. Therefore, there is no added value in varying $G_N$. 

An important thing to note about the CDF in Figure \ref{fig:CDFGN} is that there are too many galaxies with $\chi^2_{\nu}$ that is too high and also too many for which it is too low relative to the expected distribution for $\chi^2_{\nu}$. If we take the cases with high $\chi^2_{\nu}$ at face value, then the fits are formally rejected at high confidence. However, we have also made fits to the same data with dark matter halos \citep{Li2020}, and obtain CDF with the same structure for all of the halo types considered (NFW, pseudo-isothermal, Burkert, Einasto, DC14, coreNFW, and Lucky13). The same galaxies have $\chi^2_{\nu}$ that are too high for all models of any type. Taking this at face value, neither dark matter nor MOND can explain the data. Rather than conclude that all conceivable models are excluded, we infer that the uncertainties are underestimated in these cases. Similarly, the uncertainties for galaxies with very low $\chi^2_{\nu}$ have likely been overestimated. This is a common occurrence in astronomy, where there is often considerable error in the uncertainties.

The right panel of Figure \ref{fig:CDFGN} shows the different distributions for the maximum-likelihood $G_N$. The lognormal prior gives a very tight distribution around the fiducial value. It appears as a single column in log space and can be resolved by zooming-in in linear scale. This is essentially consistent with a single value of $G_N$. In contrast, the flat prior results in a wide distribution spanning $\sim$3.5 dex. This is quite similar to Figure \ref{fig:GN}, though we are fitting a different number of parameters and imposing different priors on the galactic parameters. It suggests that the wide posterior distribution is not specific to our case, but a general conclusion of flat priors on $G_N$ or g$_\dagger$. Following the same argument made in \citet{Li2018} about g$_\dagger$, we conclude that $G_N$ is indeed constant in galaxies. If insteadi, we follow the reasoning of \citet{Rodrigues2018b}, we would conclude that $G_N$ is not constant, but varies from galaxy to galaxy, just as they conclude for g$_\dagger$. In essence, they believe that there is a meaningful difference between galaxies on one side of the histogram from those on the other, while we do not.

\section{Discussion and Conclusion}

In this paper, we present an inference that Newton's constant varies as a ``reductio ad absurdum''. This highlights the dangers of degeneracies in parameter estimation from Bayesian fits when the input data are complex and their formal uncertainties cannot be taken too literally. This is motivated by the recent works of \citet{Rodrigues2018}, \citet{Chang2019}, and \citet{Marra2020}, which claim that the acceleration scale g$_\dagger$ in the empirical RAR varies from galaxy to galaxy. This conclusion is misleading; if we apply the same logic to $G_{\rm N}$, we infer that this fundamental constant varies from galaxy to galaxy. Similarly, their conclusion that MOND is ruled out at high statistical significance also applies to fits with dark matter halos \citep{Li2020}: all conceivable models are ruled out if we take the error bars literally. 

While the arguments presented here are specific to rotation-curve fits of disk galaxies, they teach us a general lesson about the application of broad priors in Bayesian analyses of astronomical data, where there is often considerable uncertainty in the uncertainties. The blind use of Bayesian statistics without properly considering the parameter degeneracy and the uncertain nature of formal errors can lead to absurd conclusions. We use $G_N$ in galaxies as an example, as it provides a straightforward comparison to the works by \citet{Rodrigues2018}, \citet{Chang2019}, and \citet{Marra2020}. The same issues could arise in other astronomical data sets.

\begin{acknowledgements}
We thank Harry Desmond for useful discussions. This work was supported in part by NASA ADAP grant 80NSSC19k0570. K.-H. C. was supported in part by the National Research Foundation of Korea (NRF) grant funded by the Korea government (MSIT) (No. NRF-2019R1F1A1062477).
\end{acknowledgements}

\bibliographystyle{aa}
\bibliography{PLi}
\end{document}